\documentclass[11pt]{article}
\usepackage[utf8]{inputenc}
\usepackage[round]{natbib}
\RequirePackage[OT1]{fontenc}
\RequirePackage{amsthm,amsmath}
\allowdisplaybreaks[4]
\usepackage{amssymb}
\usepackage{fullpage}	
\usepackage{amsthm}
\usepackage{diagbox}
\usepackage{enumerate}
\usepackage{bbm}
\usepackage{bm}
\usepackage{float}
\usepackage{makecell}
\usepackage{graphicx}
\usepackage[ruled,linesnumbered]{algorithm2e}
\usepackage{xr-hyper}
\usepackage{booktabs}
\usepackage{hyperref}[]
\hypersetup{
	colorlinks=true,
	linkcolor=blue,
	filecolor=magenta,      
	urlcolor=cyan,
	citecolor=blue,
}
\usepackage{cleveref}
\usepackage{subcaption}
\usepackage[skip=0.5ex]{caption}
\usepackage[dvipsnames, table]{xcolor}
\usepackage{multirow}
\usepackage[hmargin={1.0in, 1.0in}, vmargin={1.2in, 1.2in}]{geometry}
\usepackage{booktabs}
\usepackage{array}
\usepackage{enumitem,epsfig,natbib}
\usepackage{multirow}
\usepackage{multicol}
\newcolumntype{Z}{>{\setbox0=\hbox\bgroup}c<{\egroup}@{\hspace*{-\tabcolsep}}}
\usepackage[title]{appendix}

\newtheorem{thm}{Theorem}[section]
\newtheorem{ass}{Assumption}[section]

\newtheorem{pro}{Proposition}[section]
\newtheorem{rem}{Remark}[section]

\newtheorem{defn}{Definition}[section]

\usepackage{setspace}
\usepackage{diagbox}
\usepackage{tabu}
\usepackage{authblk}
\usepackage{xparse}
\usepackage{tikz}
\usetikzlibrary{positioning}
\usetikzlibrary{plotmarks}
\usetikzlibrary{matrix}
\usepackage{pgfplots}
\pgfplotsset{compat=1.7}
\usetikzlibrary{matrix,backgrounds, arrows.meta}
\pgfdeclarelayer{myback}
\pgfsetlayers{myback,background,main}
\usetikzlibrary{decorations.pathreplacing,angles,quotes}
\tikzset{mycolor/.style = {dashed,rounded corners,line width=1bp,color=#1}}%
\tikzset{myfillcolor/.style = {draw,fill=#1}}%
\tikzset{
	declare function={
		normcdf(\x,\m,\s)=1/(1 + exp(-0.07056*((\x-\m)/\s)^3 - 1.5976*(\x-\m)/\s));
	}
}

\newcommand{\BE}{\begin{equation}}
\newcommand{\EE}{\end{equation}}
\newcommand{\BEqn}{\begin{eqnarray*}}
\newcommand{\EEqn}{\end{eqnarray*}}

\usepackage{xr}
\makeatletter
\newcommand*{\addFileDependency}[1]{
  \typeout{(#1)}
  \@addtofilelist{#1}
  \IfFileExists{#1}{}{\typeout{No file #1.}}
}
\makeatother
\newcommand*{\myexternaldocument}[1]{%
    \externaldocument{#1}%
    \addFileDependency{#1.tex}%
    \addFileDependency{#1.aux}%
}
\myexternaldocument{supp}

\newcommand{\blind}{1}

\begin{document}
 
\if1\blind
{
 	\title{Diagnostic Checking for Wasserstein Autoregression}	

    \author[1]{Chenxiao Dai}
     \author[2]{Feiyu Jiang}
     \author[3]{Dong Li}
     \author[4]{Xiaofeng Shao}
       \affil[1]{Yau Mathematical Sciences Center, Tsinghua University}
        \affil[2]{Department of Statistics and Data Science, School of Management, Fudan University}

        \affil[3]{Department of Statistics and Data Science, Tsinghua University}
        \affil[4]{Department of Statistics and Data Science \& Department of Economics, Washington University in St. Louis}
		
	\date{}	
	\maketitle
} \fi


\begin{abstract}
Wasserstein autoregression provides a robust framework for modeling serial dependence among probability distributions, with wide-ranging applications in economics, finance, and climate science. In this paper, we develop portmanteau-type diagnostic tests for assessing the adequacy of Wasserstein autoregressive models. By defining autocorrelation functions for model errors and residuals in the Wasserstein space, we construct two related tests: one analogous to the classical McLeod type test, and the other based on the sample-splitting approach of \citet{davis2024sample}. We establish that, under mild regularity conditions, the corresponding test statistics converge in distribution to chi-square limits. Simulation studies and empirical applications demonstrate that the proposed tests effectively detect model mis-specification, offering a principled and reliable diagnostic tool for distributional time series analysis.
\end{abstract}


\section{Introduction}

Statistical modeling and analysis of object-valued time series has attracted a lot of attention lately due to the 
 increasing encounter of such data in
many modern applications. Examples include 
yearly age-at-death distributions \citep{mazzuco2015,shang2017,dubey2020frechet}, annual observations of compositions of energy sources for power generation in the U.S. \citep{zhumuller2024JoE},  and daily Pearson correlation matrices for five cryptocurrencies \citep{jiang2023two}, among others. One of the well-known challenges in dealing with such object-valued data is the lack of the usual vector/Hilbert space operations, such as scalar multiplication, addition/subtraction, and inner product, which may not be well defined and only the distance between two non-Euclidean
data objects is available. Another challenge stems from the serial dependence, which is the norm rather than the exception for many time-ordered data. 

Despite the challenge, there has been a growing literature on statistical inference for such object-valued time series. For example, change-point testing and estimation have been addressed in \cite{dubey2020frechet,jiang2023two,zhangzhushao2025}; testing for serial independence has been tackled in \cite{jiang2024testing}. On the other front, new statistical models have been developed for some specific type of object-valued time series, such as distributional time series. 
In particular, \cite{zhang2022wasserstein} proposed to use the logarithmic map  to lift the
random densities into a linear tangent space, which is a Hilbert space and the autoregressive model is imposed; also see \cite{chen2023} for related contribution along this line.
Recently, \cite{zhu2023autoregressive} developed an intrinsic approach to autoregressive modeling of one-dimensional distributional time series. The basic idea is to 
model the optimal transport map using newly defined scalar multiplication and 
addition in Wasserstein space of distributions. 
In particular, a random perturbation of the identity map was iteratively contracted/composed and applied to either the increments between consecutive distributions or to the deviations from the marginal Fr\'echet
mean, to produce autoregressive transport map (ATM) models. Subsequently \cite{jiang2022} generalized this ATM model in \cite{zhu2023autoregressive}  to the case of vector-valued distributional time series, i.e., time-evolving vectors with distributions as coordinates. Another important extension of \cite{zhu2023autoregressive} was proposed by \cite{ghodrati2023on} recently, who expanded the ATM model by allowing the shape of
the dynamics being captured by a monotone map, modulated by a contractive parameter regulating the degree of non-degeneracy of the model.

For vector-valued time series, diagnostic checking is an integral part of the model building and has been extensively studied in econometrics and time series analysis; see \cite{Li2003Diagnostic} for a book-length treatment of this topic. Generally speaking, model mis-specification can lead to biased parameter estimates, unreliable inference, and poor predictive performance, making model validation through diagnostic tools an essential step.
In this article, we study diagnostic checking of the ATM models developed in \cite{zhu2023autoregressive}, which seems lacking in the literature. Our contribution is threefold.

(a) We extend the classical definition of the autocorrelation function from vector-valued residuals to optimal transport distortion maps within the ATM framework. This enables the construction of a McLeod type portmanteau statistic \citep{mcleod1978distribution} and the derivation of its asymptotic distribution.

(b) Motivated by the recent work of \citet{davis2024sample}, who investigated sample splitting as a means to mitigate the estimation effect on the portmanteau statistic in univariate time series models, we develop an analogous sample-splitting  adjustment tailored to distributional time series under the ATM model.

(c) Through a comprehensive simulation study, we compare the performance of the classical portmanteau test and its sample-splitting counterpart, highlighting their relative size and power properties in finite samples.

We list a few commonly used notations that will appear throughout the paper; any additional notation will be introduced as needed in the main text. Let $\mathbb{R}$ denote the real line. For two functions $f, g : \mathbb{R} \to \mathbb{R}$, we write $f \circ g$ to denote their composition, that is,
$
[f \circ g](x) = f\big(g(x)\big),$ for $x \in \mathbb{R}.$
For any sequence of random
variables $\{X_n\}$, we use $\to_d$ and $\to_p$ to denote convergence in  distribution and convergence in probability,
respectively. We write $X_n=o_p(1)$  if $X_n\to_p0$ as $n\to\infty.$ 

\section{Background of optimal transport in Wasserstein space}\label{sec:back}
Let $\Omega=[\omega_s,\omega_e]$ be a closed interval on $\mathbb{R}$ and denote by $\mathcal{W}_2(\Omega)$ the space of   Borel probability measures on 
$\Omega$ with finite second moments. For  $\mu,\nu\in \mathcal{W}_2(\Omega)$, denote their corresponding distribution functions by $F_{\mu}$ and $F_{\nu}$, respectively. The 2-Wasserstein distance between $\mu$ and $\nu$ is   defined as 
$$
d^2_W(\nu,\mu)=\inf_{\gamma\in \Gamma(\nu,\mu)}\int_{\Omega}|x-y|^2\mathrm{d}\gamma(x,y),
$$
 where $\Gamma(\nu,\mu)$ denotes the set of couplings of $\nu$ and  $\mu$, i.e. Borel probability measures on $\Omega\times \Omega$ with marginal distributions $\nu$ and $\mu$.
 
 In the univariate case, if $F_{\mu}$ is absolutely continuous with respect to the Lebesgue measure, then the optimal coupling is induced by a deterministic monotone transport map
\[
T(x)= F_{\nu}^{-1}\!\big(F_{\mu}(x)\big),
\]
in which $\nu=T\#\mu$ is the pushforward measure of $\mu$ by $T$, and as a consequence,
\[
d_W^2(\mu,\nu) = \int_{\Omega} |T(x) - x|^2 \, \mathrm{d}\mu(x)
= \int_0^1 \big| F_{\mu}^{-1}(u) - F_{\nu}^{-1}(u) \big|^2 \, \mathrm{d}u.
\]
In this paper, we restrict our attention to  the space of monotone transport maps, i.e. 
\[
\mathcal{T} = \{\, T:\Omega \to \Omega \mid T \text{ is nondecreasing and } T(\omega_s)=\omega_s,\, T(\omega_e)=\omega_e \,\}. 
\]
Therefore, since both  $\omega_s$ and $\omega_e$ are bounded, all transports in $\mathcal{T}$ (and their inverse transports) are bounded.

The ATM model is built upon a class of iterated random systems of transport maps known as ``contractive compositions".
A key ingredient is the notion of an $\alpha$-contraction of a transport map.
In particular, for a parameter $\alpha \in [-1,1]$ and any optimal transport map $T\in\mathcal{T}$, the $\alpha$-contraction is defined by the operator   $T\mapsto [\alpha \odot T],$ where
$$
[\alpha \odot T](x)= \begin{cases}x+\alpha(T(x)-x), & 0<\alpha \leq 1, \\ x, & \alpha=0, \\ x+\alpha\left(x-T^{-1}(x)\right), & -1 \leq \alpha<0.\end{cases}
$$

Intuitively speaking, the operator $[\alpha \odot T]$ attenuates the deformation induced by $T$. When $0<\alpha\leq 1,$  the map moves each point $x$ only a fraction 
$\alpha$ of the way along the transport path from $x$ to $T(x)$, thereby ``contracting" the action of 
 $T$ toward the identity. When $\alpha\leq 0$, the movement is partly reversed, shifting 
$x$ in the opposite direction of the original transport and producing a reflected contraction.
In short, the $\alpha$-contraction 
generates a controlled and smoothly scaled version of the transformation induced by $T$, forming the building block of the ATM dynamics.

\section{ATM model and   estimation}\label{sec:atm}
This section provides concise background information on ATM models in \cite{zhu2023autoregressive}. We begin by introducing the model setup, then derive the asymptotic distribution of the estimated parameters under the assumption that the underlying distributional data are fully observed.

\subsection{Autoregressive Transport Map model}
Let $\{\mu_i\}_{i=1}^n \subset \mathcal{W}_2(\Omega)$ be a sequence of stationary probability measures on $\Omega$, with the Wasserstein barycenter defined as 
$$
\mu_F=\arg\min_{\nu\in \mathcal{W}_2(\Omega)}\mathbb{E}d^2_W(\nu,\mu_i).
$$
In this paper, we focus on the associated optimal transport maps of the form
\begin{flalign}
    T_i = F^{-1}_{\mu_i} \circ F_{\mu_F}, \quad i = 1, \dots, n,
\end{flalign}
which transport the barycenter measure $\mu_F$ to each $\mu_i$. 
{\footnote{Alternatively, one may model the incremental transports by considering
$
T_i = F^{-1}_{\mu_i} \circ F_{\mu_{i-1}},
$
which captures the incremental evolution of the distributions over time.}}

Following \citet{zhu2023autoregressive}, the first-order autoregressive transport map model (ATM(1)) is \begin{equation}\label{atm_model}
T_i=T_{\epsilon_i}\circ[\alpha  \odot T_{i-1}],
\end{equation}
where $\alpha\in(-1,1)$ is the model parameter, $T_{\epsilon_i}$ is a sequence of  i.i.d.\ innovation distortion maps satisfying $\mathbb{E}T_{\epsilon_i}(x)=x$ for all $x\in\Omega$ almost surely.
 
The key idea of the ATM model is to mimic the classical autoregressive dynamics within the Wasserstein space by replacing vector addition and scalar multiplication with their transport map counterparts---composition and $\alpha$-contraction, respectively. 
Specifically, the map $T_i$ evolves by composing the $\alpha$-contracted previous map $[\alpha \odot T_{i-1}]$ with a random innovation map $T_{\epsilon_i}$, thereby capturing temporal dependence in the evolution of probability measures $\{\mu_i\}$. 
Under appropriate contraction and moment conditions, the sequence $\{T_i\}$ admits a unique stationary distribution in $\mathcal{T}$, and the induced process $\{\mu_i\}$ is stationary in $\mathcal{W}_2(\Omega)$.

For any transport map $t\in\mathcal{T}$, and $\alpha \in(-1,1)$, define the composite operator
$$\phi_{\epsilon}(t)= T_\epsilon\circ [\alpha \odot t], \quad \text{and}\quad 
\widetilde{\phi}_{i,m}(t)=\phi_{\epsilon_i}\circ\phi_{\epsilon_{i-1}}\circ \cdots\circ \phi_{\epsilon_{i-m+1}}(t).
$$

\begin{ass}\label{ass_contraction}
The sequence $\{T_{\epsilon_i}\}$ consists of i.i.d. distortion maps satisfying $\mathbb{E}\{T_{\epsilon_i}(x)\}=x$ almost surely for all $x\in \Omega$. Moreover, there exists constants $C>0$, and $r\in(0,1)$, such that for all $t, t'\in\mathcal{T}$ and all $m\in\mathbb{N}$, 
    $$
\mathbb{E}\left[d_1\left(\widetilde{\phi}_{i, m}\left(t\right), \widetilde{\phi}_{i, m}(t')\right)\right] \leq C r^m d_1\left(t,t'\right),
$$
where $d_1(t,s)=\int_\Omega\|t(x)-s(x)\|\mathrm{d}x$ is the $L_1$ distance for $t,s\in \mathcal{T}$.
\end{ass}
\Cref{ass_contraction} imposes a moment contraction condition \citep{wu2004limit} on the ATM model, which guarantees the existence and uniqueness of a stationary solution to the model dynamics in \eqref{atm_model}. 
Similar conditions have been adopted in \citet{zhu2023autoregressive} and \citet{ghodrati2023on};  we employ the $L_1$ distance metric here, primarily to simplify the derivation of the estimation theory.

The following proposition is a direct consequence of  Theorem 1 in \cite{zhu2023autoregressive}. 
\begin{pro}
Under \Cref{ass_contraction}, for any $T_0\in\mathcal{T}$, $\widetilde{T}_i:=\lim_{m\to\infty}\widetilde{\phi}_{i,m}(T_0)\in\mathcal{T}$ exists almost surely and does not depend on $T_0$. Moreover, $\widetilde{T}_i$ is a unique and stationary solution to the model \eqref{atm_model}.
\end{pro}

\subsection{Estimation}
Suppose the true  model \eqref{atm_model} is parameterized by $\alpha_0\in(-1,1)$, then it is  the unique minimizer of $$
l(\alpha)=\mathbb{E}\int_\Omega \left\{T_{i+1}(x)-[\alpha\odot T_i](x)\right\}^2\mathrm{d}x.
$$
In other words, we have that 
$$
\alpha_0=\left\{\begin{array}{cc}
\frac{\int_{\Omega} \mathbb{E}\left[\left(T_{i+1}(x)-x\right)\left(T_i(x)-x\right)\right] \mathrm{d} x}{\int_{\Omega} \mathbb{E}\left[\left(T_i(x)-x\right)^2\right]\mathrm{d} x}, & \text { if } 0\leq \alpha_0 < 1 \\
\frac{\int_{\Omega} \mathbb{E}\left[\left(T_{i+1}(x)-x\right)\left(x-T_i^{-1}(x)\right)\right] \mathrm{d} x}{\int_{\Omega} \mathbb{E}\left[\left(x-T_i^{-1}(x)\right)^2\right]\mathrm{d} x}, & \text { if } -1<\alpha_0\leq 0.
\end{array}\right.
$$
Therefore, following \cite{zhu2023autoregressive}, we estimate $\alpha_0$ by the least square type estimator,
\begin{equation}\label{alpha_est}
\widehat{\alpha}_n=\widehat{\alpha}_+\mathbb{I}{\{l_+(\widehat{\alpha}_+)\leq l_-(\widehat{\alpha}_-)\}}+\widehat{\alpha}_-\mathbb{I}{\{l_+(\widehat{\alpha}_+)>l_-(\widehat{\alpha}_-)\}},
\end{equation}
where $\widehat{\alpha}_{+}=\operatorname{argmin}_\alpha l_{+}(\alpha), \widehat{\alpha}_{-}=\operatorname{argmin}_\alpha l_{-}(\alpha)$, and 
\begin{flalign*}
    l_+(\alpha)=&\sum_{i=2}^n\int_{\Omega}\left\{T_i(x)-x-\alpha(T_{i-1}(x)-x)\right\}^2\mathrm{d}x,\\l_-(\alpha)=&\sum_{i=2}^n\int_{\Omega}\left\{T_i(x)-x-\alpha(x-T_{i-1}^{-1}(x))\right\}^2\mathrm{d}x.
\end{flalign*}

Note that the estimation step involves comparing two different objective functions corresponding to positive and negative values of the contraction parameter $\alpha$, making the theoretical analysis substantially more involved than that for classical least-squares estimators. 
To this end, we employ the concentration inequality developed in \citet{wu2004limit} and show that, the estimation of the sign of $\alpha_0$ is super-consistent, 
thereby ensuring that the estimator $\widehat{\alpha}_n$ correctly identifies the contraction regime with probability approaching one. 
This is crucial for establishing the subsequent asymptotic linearity and normality of $\widehat{\alpha}_n$. 

\begin{thm}\label{thm_alpha}
    Let $\{T_i\}_{i=1}^{n}$ be a sequence of observed optimal transportation maps generated by model \eqref{atm_model} satisfying $\mathbb{E}\int_{\Omega}(T_i(x)-x)^2\mathrm{d}x\geq \underline{c}$ for some constant $\underline{c}>0$. Let $\widehat{\alpha}_n$ be defined in \eqref{alpha_est}.  Under \Cref{ass_contraction}, as $n\to\infty$, 
    $$
    \sqrt{n}(\widehat{\alpha}_n-\alpha_0)=\frac{1}{\sqrt{n}}\sum_{i=1}^{n-1} m_i(\alpha_0)+o_p(1),
    $$
where for $\Xi=\mathbb{E}\int_{\Omega}[T_i(x)-x]^2\mathrm{d}x$ and $\widetilde{\Xi}=\mathbb{E}\int_{\Omega}[T_i^{-1}(x)-x]^2\mathrm{d}x$, 
$$m_i(\alpha)= 
\begin{cases}
   \Xi^{-1}\int_{\Omega}\{T_{i+1}(x)-[\alpha\odot T_{i}](x)\}(T_{i}(x)-x)\mathrm{d}x & \alpha\geq 0\\
    \widetilde{\Xi}^{-1} \int_{\Omega}\{T_{i+1}(x)-[\alpha\odot T_{i}](x)\}(x-T_i^{-1}(x))\mathrm{d}x, &\alpha<0.
\end{cases}
$$
Furthermore, for the   natural filtration $\mathcal{F}_i=\sigma(T_j, j\leq i)$,  $\{m_i(\alpha_0);\mathcal{F}_{i}\}$ forms a martingale difference sequence.
\end{thm}
The lower bound condition  $\mathbb{E}\int_{\Omega}(T_i(x)-x)^2\mathrm{d}x\geq \underline{c}$ is imposed solely to exclude degenerate cases of the transport map of $T_i(x)$. As otherwise, $T_i(x)=x$ for $x\in\Omega$ a.s., which corresponds to a trivial static setting.  By the boundedness of $T_i\in\mathcal{T}$, the classical central limit theorem for martingales \citep{hall2014martingale} and \Cref{thm_alpha} together imply that $\widehat{\alpha}_n$ is asymptotically normal, i.e. 
$$
 \sqrt{n}(\widehat{\alpha}_n-\alpha_0)\to_d\mathcal{N}(0,\mathbb{E}[m_i(\alpha_0)]^2),\quad \text{as}~~ n\to\infty.
$$

In this paper, we assume the data are fully observed,  whereas \citet{zhu2023autoregressive} allow for discretely observed distributions, where the transport maps can be estimated from empirical grid observations. Compared with the convergence rate result in \citet{zhu2023autoregressive}, \Cref{thm_alpha} further establishes the asymptotic distribution of the estimator $\widehat{\alpha}_n$, thereby enabling valid statistical inference for the contraction parameter $\alpha$. A similar fully observed setting is also considered in \citet{ghodrati2023on}, where an additional ``intercept'' term is introduced to capture the mean shape of the transport maps. 
However, extending the asymptotic normality result to the case of estimated or partially observed maps is considerably more challenging, as the estimation error of the empirical transport maps propagates nonlinearly through the composition and contraction operations. We leave them for future research.

\section{Diagnostic Tests}\label{sec:test}
In this section, we develop diagnostic procedures to evaluate the adequacy of the ATM(1) model. 
We first formalize the definitions of residual transport maps and their autocorrelation functions,  which serve as the primary tools for assessing model fit. 
Next, we introduce  McLeod \citep{mcleod1978distribution} type portmanteau tests to detect serial dependence in the residuals, thereby identifying potential model mis-specification.  Finally, we propose a sample–splitting–based adjustment \citep{davis2024sample} to the portmanteau tests, which helps mitigate the impact of parameter estimation and improves the reliability of inference.

\subsection{Residuals and auto-correlation functions}
To assess the adequacy of the fitted ATM model, we define residual maps that quantify the deviation between the observed and fitted transport maps.  By model \eqref{atm_model}, with $\{T_i\}$ and model parameter $\alpha_0$, we can recover the distortion map as $$
 T_i\circ[\alpha_0 \odot T_{i-1}]^{-1}={T}_{\epsilon_i}\circ[\alpha_0 \odot T_{i-1}]\circ[\alpha_0 \odot T_{i-1}]^{-1}={T}_{\epsilon_i}.
$$ 
Therefore, given the estimated contraction parameter $\widehat{\alpha}_n$, the residual is naturally defined as
\begin{flalign}\label{eq:res}
\widehat{T}_{\epsilon_i} =  T_i \circ [\widehat{\alpha}_n \odot T_{i-1}]^{-1}, \quad i = 2, \dots, n.
\end{flalign}
Under the correctly specified model, the estimated innovation maps $\{\widehat{T}_{\epsilon_i}\}$ should mimic the behavior of the true innovations $\{T_{\epsilon_i}\}$. Similar to the classical Euclidean autoregressive model, the temporal dependence of $\{T_{\epsilon_i}\}$ can be assessed via the autocorrelation function (ACF).  \begin{defn}\label{defn_acf}
The theoretical ACF at lag $k\geq 1$ for the distortion map process $\{T_{\epsilon_i}\}$ is given by 
    $$
\rho(k)= \frac{\int_{\Omega} \mathbb{E}\left[\left(T_{\epsilon_{i+k}}(x)-x\right)\left(T_{\epsilon_{i}}(x)-x\right)\right]\mathrm{d} x}{\int_{\Omega} \mathbb{E}\left[\left(T_{\epsilon_{i}}(x)-x\right)^2\right] \mathrm{d} x}.
$$
\end{defn} 
Here, the definition of ACF  projects the distortion maps onto  $L^2(\Omega)$ space and assesses temporal dependence by integrating the pointwise autocorrelations of $T_{\epsilon_i}(x)$ over $\Omega$. While possibly the pointwise correlations may cancel out during integration, the resulting global autocorrelation measure still provides a meaningful overall measure of temporal dependence. 
 It serves as the basis for constructing the portmanteau statistic that follows.

\subsection{Portmanteau Test}
Given estimated residual transports $\widehat{T}_{\epsilon_2},\cdots,\widehat{T}_{\epsilon_n}$ defined in \eqref{eq:res}, the corresponding sample ACF at lag $k$ is
\begin{equation}\label{ACF}
  \widehat{\rho}_n(k)= \frac{\int_{\Omega} \sum_{i=1}^{n-k}\left[\left(\widehat{T}_{\epsilon_{i+k}}(x)-x\right)\left(\widehat{T}_{\epsilon_{i}}(x)-x\right)\right] \mathrm{d} x}{\int_{\Omega} \sum_{i=1}^n\left[\left(\widehat{T}_{\epsilon_{i}}(x)-x\right)^2\right] \mathrm{d} x}.  
\end{equation}

If $\widehat{\alpha}_n$ consistently estimates the true parameter $\alpha$, then the estimated residual maps $\widehat{T}_{\epsilon_i}$ should approximate the true distortion maps $T_{\epsilon_i}$, thereby preserving their statistical properties.
Hence,  under the correctly specified model, the sample autocorrelations $\widehat{\rho}_n(k)$ in \eqref{ACF} are expected to be close to zero for all $k \geq 1$, reflecting the absence of serial dependence among the residual transports. 
Significant deviations from zero would indicate potential model mis-specification, motivating the use of portmanteau-type tests. 

While similar ideas have been widely used in the classical Euclidean time-series literature—see e.g. \citet{ljung1978measure} and  \citet{mcleod1978distribution}, the extension to the Wasserstein setting requires substantial additional work due to the nonlinear structure of transport maps.
 Before presenting the asymptotic distribution of $\widehat{\rho}_n(k)$, we introduce additional notations that will be used in the subsequent analysis.

Define $G_i(\alpha,x)=T_i\circ [{\alpha} \odot T_{i-1}]^{-1}(x)$, and let $g_i(\alpha,x)$  be the derivative of $G_i(\alpha,x)$ with respect to $\alpha$, where by  equation (31) in \cite{ghodrati2023on},
\begin{flalign}\label{eq:derivative}
\begin{split}
    g_i(\alpha,x)= &\frac{\partial G_i(\alpha,x)}{\partial \alpha} =T_i^{\prime}\left(z\right) \times 
   \begin{cases}\left(z-T_{i-1}\left(z\right)\right) \times\left.\frac{1}{\alpha\left(T_{i-1}^{\prime}\left(z\right)-1\right)+1}\right|_{z=\left[\alpha \odot T_{i-1}\right]^{-1}(x)}, & 0\leq \alpha <1 ,
   \\
   \left(T_{i-1}^{-1}\left(z\right)-z\right) \times\left.\frac{1}{\alpha\left(1-\left(T_{i-1}^{-1}\right)^{\prime}\left(z\right)\right)+1}\right|_{z=\left[\alpha \odot T_{i-1}\right]^{-1}(x)}, & -1  <\alpha\leq 0.
\end{cases}   
\end{split}
\end{flalign}
Fix an integer $K\geq 1$, for $1\leq k\leq K,$ define  $M_1=(\mu_{11},\cdots, \mu_{1K})^{\top},$ $M_2=(\mu_{21},\cdots, \mu_{2K})^{\top},$ where
$$\mu_{1k}=\mathbb{E}\left[\int_{\Omega}(T_{\epsilon_0}(x)-x)g_k(\alpha_0,x)\mathrm{d}x\right],\quad \mu_{2k}=\mathbb{E}\left[\int_{\Omega} \left[\left(T_{\epsilon_0}(x)-x\right)\left(T_{\epsilon_{k}}(x)-x\right)\right] \mathrm{d}x\cdot m_k(\alpha_0)\right].$$
Furthermore,  let 
$$\sigma_1^2=\mathbb{E}\left\{\int_{\Omega} [T_{\epsilon_i}(x)-x]^2\mathrm{d}x \right\}, \quad \sigma_2^4=\mathbb{E}\left\{\int_{\Omega} [T_{\epsilon_1}(x)-x] [T_{\epsilon_2}(x)-x] \mathrm{d}x \right\}^2.$$

\begin{ass}\label{ass_lowerbound}
There exists a constant $c>0$ such that $\sup_{\alpha\in [-1,1]}\mathbb{E}\int_{\Omega}[g(\alpha,x)]^2\mathrm{dx}\leq c<\infty$.
\end{ass} 
\Cref{ass_lowerbound}  imposes a mild second moment condition on the derivative function. 
\begin{thm}\label{thm_port}
Under the correct model specification \eqref{atm_model} with parameter $\alpha_0\in(-1,1)$, let $\{\widehat{T}_{\epsilon_i}\}_{i=2}^n$ be the residual maps defined in \eqref{eq:res}. For any fixed integer $K\geq 1$,   under \Cref{ass_contraction} and \ref{ass_lowerbound}, as $n\to\infty$, we have that 
 $$\sqrt{n}\widehat{\rho}_n^\top=\sqrt{n} \left(\widehat{\rho}_{n}(1), \cdots, \widehat{\rho}_{n}(K)\right)^\top\to_d\mathcal{N}(\mathbf{0}, \Sigma_K),$$
where $$\Sigma_K:=\sigma_1^{-4}\left\{\sigma_2^4I_K+(M_1M_2'+M_2M_1'+M_1\mathbb{E}[m_i^2(\alpha_0)]M_1^{\top})\right\}.$$
\end{thm}
\Cref{thm_port} establishes the joint asymptotic normality of the sample autocorrelations of the residual transport maps. 
The presence of $M_1$ and $M_2$ reflects the nonlinear dependence of the residuals on the estimated contraction parameter $\alpha_0$ via the transport-map derivatives $g_i$.  
The additional term $M_1\,\mathbb{E}[m_i^2(\alpha_0)]\, M_1^\top$ captures the accumulated effect of parameter estimation on the limiting distribution in \Cref{thm_alpha}, analogous to the correction introduced by \citet{mcleod1978distribution,li1981distribution} for finite-sample adjustment to the classical Ljung-Box test in \cite{ljung1978measure}.  

In practice, we estimate the covariance matrix ${\Sigma}_K$ by its sample counterpart $$\widehat{\Sigma}_K=\widehat{\sigma}_1^{-4}\left\{\widehat{\sigma}_2^4I_K+(\widehat M_1\widehat M_2'+\widehat M_2\widehat M_1'+\widehat M_1\frac{1}{n}\sum_{i=1}^n[\widehat{m}_i^2(\widehat{\alpha}_n)]\widehat M_1^{\top})\right\},$$ where for 
 $\widehat{M}_1=(\widehat\mu_{11},\cdots, \widehat\mu_{1K})^{\top},$ $\widehat M_2=(\widehat\mu_{21},\cdots, \widehat\mu_{2K})^{\top},$ 
\begin{flalign*}
\widehat\mu_{1k}=&\frac{1}{n-k}\sum_{i=1}^{n-k}\left[\int_{\Omega}(\widehat{T}_{\epsilon_i}(x)-x)g_{i+k}(\widehat{\alpha}_n,x)\mathrm{d}x\right],\\ \widehat\mu_{2k}=&\frac{1}{n-k}\sum_{i=1}^{n-k}\left[\int_{\Omega} \left[\left(\widehat{T}_{\epsilon_i}(x)-x\right)\left(\widehat{T}_{\epsilon_{i+k}}(x)-x\right)\right] \mathrm{d}x\cdot \widehat{m}_{i+k}(\widehat{\alpha}_n)\right],
\end{flalign*}
and $$\widehat{m}_i(\alpha)= 
\begin{cases}
   \left(\frac{1}{n}\sum_{i=1}^n[T_i(x)-x]^2\mathrm{d}x\right)^{-1}\int_{\Omega}\{T_{i+1}(x)-[\alpha\odot T_{i}](x)\}(T_{i}(x)-x)\mathrm{d}x, & \alpha\geq 0,\\\\
    \left(\frac{1}{n}\sum_{i=1}^n[T_i^{-1}(x)-x]^2\mathrm{d}x\right)^{-1} \int_{\Omega}\{T_{i+1}(x)-[\alpha\odot T_{i}](x)\}(x-T_i^{-1}(x))\mathrm{d}x, &\alpha<0,
\end{cases}
$$
\begin{equation}\label{sigma_est}
    \widehat \sigma_1^2=\frac{1}{n}\sum_{i=1}^n \int_{\Omega} [\widehat{T}_{\epsilon_i}(x)-x]^2\mathrm{d}x, \quad \widehat{\sigma}_2^4=\left\{\frac{1}{n}\sum_{i=1}^{n-1} \int_{\Omega} [\widehat{T}_{\epsilon_i}(x)-x] [\widehat{T}_{\epsilon_{i+1}}(x)-x] \mathrm{d}x \right\}^2.
\end{equation}

Then, the McLeod type portmanteau test statistic can be formulated as
\[
Q_n = n\,\widehat{\rho}_n^\top \widehat{\Sigma}_K^{-1}\,\widehat{\rho}_n,
\]
and we reject the null hypothesis of correct model specification if 
\begin{equation}\label{test_1}
    Q_n > \chi^2_{K}(1-\beta),
\end{equation}
the $(1-\beta)$ upper quantile of the chi-square distribution with $K$ degrees of freedom.

\begin{rem}
In classical univariate ARMA models, portmanteau test statistics are typically constructed as
\begin{equation}\label{eq:Qstar}
    Q_n^* = n\sum_{k=1}^K w_n(k)\,\widehat{\rho}_n^2(k),
\end{equation}
where $w_n(k)=1$ corresponds to the Box--Pierce statistic \citep{box1970distribution}, and
$w_n(k)=(n+2)/(n-k)$ yields the Ljung--Box statistic \citep{ljung1978measure}. To improve the
finite-sample approximation, \citet{li1981distribution} further proposed a modified statistic of
the form
\begin{equation}\label{limcleod}
    Q_n^{\dagger} = Q_n^* + \frac{K(K+1)}{2n}.
\end{equation}
Then, the null hypothesis is rejected when $Q_n^*$ (or $Q_n^{\dagger}$) exceeds the
$(1-\alpha)$ quantile of a chi-squared distribution with $K-m$ degrees of freedom, where $m$
denotes the total number of estimated ARMA parameters. This chi-square reference relies on an
\emph{increasing-lag} asymptotic framework, that is, $K = K(n) \to \infty$ as $n \to \infty$ with
$
    n \gg K \gg 1,
$
see Chapter~2 in \citet{Li2003Diagnostic} for a detailed discussion. In this diverging-$K$ regime,
the effective degrees of freedom reduce from $K$ to $K-m$ because the covariance matrix of residual
autocorrelations becomes increasingly singular  (and thus not directly invertible) as $K$
grows. Consequently, the classical portmanteau tests of the form \eqref{eq:Qstar} or \eqref{limcleod} can be implemented without inverting the covariance matrix, and instead rely on weighting schemes or small-sample
corrections to improve the chi-square approximation.

However, in this paper (and arguably in typical empirical applications), we treat $K$ as a fixed
integer. In this fixed-$K$ regime, the asymptotic covariance matrix of the residual ACF is
non-singular, and following the recommendations of \citet{mcleod1978distribution},
\citet{ansley1979finite}, and \citet{newbold1980equivalence}, we explicitly estimate this
covariance matrix to account for the parameter estimation effect. This kind of treatment has been also used in diagnostic checks for nonlinear time series models, see e.g. \cite{li1992asymptotic} and \cite{li1994squared}.

\end{rem}

\subsection{Sample Splitting Approach}

Despite the finite-sample correction for estimation effects provided in \Cref{thm_port}, estimating quantities such as $M_1$, $M_2$, and $\mathbb{E}[m_i^2(\alpha_0)]$ may be challenging in practice, as they involve expectations of nonlinear functionals of transport maps and their derivatives, especially in moderate sample sizes. 
This motivates the development of alternative diagnostic procedures that avoid direct estimation of these higher-order terms while still accounting for the impact of parameter estimation.

Sample splitting has recently emerged as an effective tool for time-series inference, with theoretical guarantees established in several distinct settings; see, for example, \citet{Lunde19}, \citet{Chang21}, \citet{zhangshao2023}, and \citet{gaowangshao2023}. 
Of particular relevance is the work of \citet{davis2024sample}, who employed sample splitting to remove estimation effects in portmanteau tests applied to residuals of parametric time-series models, such as ARMA and GARCH models.
Their methodology estimates the model parameters using only the first $f_n$ observations, and then computes the autocorrelation function (ACF) of the residuals based solely on the last $l_n$ observations. 
They show that as long as the relative overlap between the two subsamples converges to $1/2$, the resulting ACF-based tests retain the same limiting distribution as if the underlying residuals were i.i.d. 
Consequently, the classical portmanteau test admits its convenient chi-square limiting distribution when sample splitting with appropriately chosen $(f_n,l_n)$ is employed—for instance, $f_n=n/2$ together with $l_n=n$.

Here, we extend the idea of \cite{davis2024sample} to the non-Euclidean setting of distributional data.  In particular, let $\widehat{\alpha}_{f_n}$ be the sample-splitting analogue of \eqref{alpha_est} using first $f_n$ observations $\{T_i\}_{i=1}^{f_n}$, and the residual be
\begin{equation}\label{eq:res_split}
\widehat{T}_{\epsilon_i}^{f_n}=T_i \circ [\widehat{\alpha}_{f_n} \odot T_{i-1}]^{-1}, \quad i = n-l_n+1, \dots, n, 
\end{equation}
which reduces to the standard case in \eqref{eq:res} if $f_n=l_n=n$.
Similarly, for $1\leq k\leq K$, we define the lag-$k$ sample autocorrelation function as 
$$
\widehat{\rho}_{l_n}(k)=\frac{\int_{\Omega} \sum_{i=n-l_n+1}^{n-k}\left[\left(\widehat{T}_{\epsilon_{i+k}}^{f_n}(x)-x\right)\left(\widehat{T}_{\epsilon_i}^{f_n}(x)-x\right)\right] \mathrm{d} x}{\int_{\Omega} \sum_{i=n-l_n+1}^n\left[\left(\widehat{T}_{\epsilon_i}^{f_n}(x)-x\right)^2\right] \mathrm{d} x}.  
$$

The following result generalizes \Cref{thm_port} to the sample-splitting case. 

\begin{thm}\label{thm_split}
   Under the correct model specification \eqref{atm_model} with parameter $\alpha_0\in(-1,1)$, let $\{\widehat{T}^{f_n}_{\epsilon_i}\}_{i=n-l_n+1}^n$ be the residual maps defined in \eqref{eq:res_split}. Suppose the split sequence $(f_n,l_n)$ satisfies that $\min\{f_n,l_n\}\to \infty$ as $n\to\infty$. Let 
\[
k_{ra}=\lim_{n\to\infty}\frac{l_n}{f_n}<\infty
\quad\text{and}\quad
k_{ov}=\lim_{n\to\infty}\frac{\max\{0,\,f_n+l_n-n\}}{f_n}<\infty,
\]
denote the limiting ratio and limiting overlap coefficient, respectively.  Then, for any fixed integer $K\geq 1$,   under \Cref{ass_contraction} and \ref{ass_lowerbound}, as $n\to\infty$, we have that 
    $$\sqrt{l_n}\widehat{\rho}_{l_n}^{\top}=\sqrt{l_n} (\widehat{\rho}_{l_n}(1), \cdots, \widehat{\rho}_{l_n}(K))^\top\to_d\mathcal{N}(\mathbf{0}, \Sigma_K^S),$$
    where    $$\Sigma_K^S:=\sigma_1^{-4}\left\{\sigma_2^4I_K+(k_{ov}M_1M_2^{\top}+k_{ov}M_2M_1^{\top}+k_{ra}M_1\mathbb{E}[m_i^2(\alpha_0)]M_1^{\top})\right\}.$$
    Furthermore, if  \begin{equation}\label{split_condition}
        M_2=-M_1\mathbb{E}[m_i^2(\alpha_0)],
    \end{equation} then $\Sigma_K^S$ simplifies to 
$$\Sigma_K^S=\sigma_1^{-4}\left\{\sigma_2^4I_K+(k_{ra}-2k_{ov})M_1\mathbb{E}[m_i^2(\alpha_0)]M_1^{\top}\right\}.$$
\end{thm}

The result in \Cref{thm_split} mirrors that of \Cref{thm_port}, with the key difference being the appearance of the split sequence $(f_n,l_n)$ and its limiting behavior.
In particular, when $f_n = l_n = n$, \Cref{thm_split} reduces exactly to \Cref{thm_port}.

In \Cref{thm_split}, the condition \eqref{split_condition} that $M_2=-M_1\mathbb{E}[m_i^2(\alpha_0)]$ is motivated by the findings of \citet{davis2024sample}, who showed that it holds for classical ARMA and GARCH models. 
Establishing this identity analytically for the ATM model is substantially more difficult, as it involves functional derivatives such as those appeared in \eqref{eq:derivative}. 
To assess its validity in the ATM setting, we therefore conduct a numerical investigation under the ATM(1) specification.

 We consider three different forms of the innovation map $T_{\epsilon}(x)$: trigonometric, power, and polynomial. 
\begin{align*}
e_1(x) &= x + \frac{\sin(Y_1\pi x)}{|Y_1|\pi},\\
e_2(x) &= x + Y_2(x^2 - x),\\
e_3(x) &= x + Y_3x^2(1 - x)^2,
\end{align*}
where $Y_1$ is a random variable generated from a discrete uniform distribution over $\{\pm5,\pm6,\cdots,\pm15\}$, and $Y_2, Y_3$ are random variables generated from a uniform distribution over $[-1,1]$. 

We numerically evaluate the discrepancy between $M_2$ and $-M_1\mathbb{E}[m_i^2(\alpha_0)]$  by varying $\alpha_0\in\{-0.4,-0.2,0.2,0.5\}$, fixing $K=12$. Specifically, we generate $n=5000$
observations from model \eqref{atm_model}, and compute the empirical estimates of both $M_2$ and $-M_1\mathbb{E}[m_i^2(\alpha_0)]$.  Table \ref{tab:distances_results_mean_std}  reports the mean $L_1$ and $L_2$ distances between the two quantities over 100 repetitions, with standard deviations shown in parentheses. It can be seen that all reported discrepancy measures across the three innovation families are uniformly small, with mean values consistently close to zero.

\begin{table}[htbp]
\centering
\caption{Numerical validation of condition \eqref{split_condition}.}
\label{tab:distances_results_mean_std}
\footnotesize 
\setlength{\tabcolsep}{3pt} 
\begin{tabular}{cccccc}
\toprule
\multirow{2}{*}{Function} & \multirow{2}{*}{Difference} & \multicolumn{4}{c}{Mean ± (Standard Deviation)} \\
\cmidrule(r){3-6}
 & & $\alpha=0.5$ & $\alpha=0.2$ & $\alpha=-0.2$ & $\alpha=-0.4$ \\
\midrule
\multirow{2}{*}{$e_1(x)$} 
 & $L^1$ & $ 0.0003\pm(0.0006) $ & $ 0.0003\pm(0.0005) $ & $ 0.0102\pm(0.0036) $ & $ 0.0425\pm(0.1300) $ \\
 & $L^2$ & $ 0.0001\pm(0.0002) $ & $ 0.0001\pm(0.0001) $ & $ 0.0029\pm(0.0003) $ & $ 0.0123\pm(0.0108) $ \\
\cmidrule{1-6}
\multirow{2}{*}{$e_2(x)$}
 & $L^1$ & $ 0.0170\pm(0.0155) $ & $ 0.0142\pm(0.0131) $ & $ 0.0321\pm(0.0004) $ & $ 0.0752\pm(0.0540) $ \\
 & $L^2$ & $ 0.0050\pm(0.0045) $ & $ 0.0042\pm(0.0038) $ & $ 0.0099\pm(0.0000) $ & $ 0.0223\pm(0.0045) $ \\
\cmidrule{1-6}
\multirow{2}{*}{$e_3(x)$}
 & $L^1$ & $ 0.0035\pm(0.0033) $ & $ 0.0028\pm(0.0026) $ & $ 0.0013\pm(0.0000) $ & $ 0.0014\pm(0.0000) $ \\
 & $L^2$ & $ 0.0010\pm(0.0009) $ & $ 0.0008\pm(0.0008) $ & $ 0.0004\pm(0.0000) $ & $ 0.0004\pm(0.0000) $ \\
\bottomrule
\end{tabular}
\end{table}

A direct consequence of condition \eqref{split_condition} is its practical implication: when the splitting sequence satisfies $k_{ra}=2k_{ov}$, the resulting covariance matrix 
$\Sigma_K^S$
 becomes nearly pivotal, up to the nuisance scale parameters 
$\sigma_1$ and $\sigma_2$, both of which can be readily estimated from empirical residuals using \eqref{sigma_est}. 

In particular, if we  choose $f_n=n/2$ and $l_n=n$, that is,  we use the first half of the data to form an estimator, and then use the full data to compute residuals,  we can easily obtain that $\Sigma_K^S=\sigma_2^4I_K/\sigma_1^4$. Consequently,  $\Sigma_K^S$ can be consistently estimated by $
\widehat{\Sigma}_K^S={\widehat{\sigma}_2^4} I_K/{\widehat{\sigma}_1^4}.
$
Therefore, the sample-splitting based test statistic is given by
\begin{equation}\label{test_2}
   Q_n^S= n\widehat{\rho}_{n}^{\top}(\widehat{\Sigma}_K^S)^{-1}\widehat{\rho}_{n} = \frac{\widehat{\sigma}_1^4}{\widehat{\sigma}_2^4}n\widehat{\rho}_{n}^{\top}\widehat{\rho}_{n}, 
\end{equation}
and   we reject the null hypothesis of correct model specification if 
$
Q_n^S> \chi^2_{K}(1-\beta),$ with significance level $\beta\in(0,1).$

\section{Simulation Results}\label{sec:simu}
This section investigates the finite-sample performance of the proposed portmanteau tests. 
Following \citet{ghodrati2023on}, the distortion maps are generated according to 
\begin{equation}\label{eq:distortion}
    T_{\epsilon_i}(x)=x+\frac{\sin (Y\pi x)}{|Y|\pi},
\end{equation}
where $Y$ is a random variable generated from a discrete uniform distribution over $\{\pm5,\pm6,\cdots,\pm15\}$.  

To evaluate size performance, we simulate the ATM(1) model under various specifications with 
$\alpha_0 \in \{-0.4, -0.2, 0.2, 0.5\}$, and sample sizes $n \in \{100, 200, 400\}$. 
Table \ref{tab_size} reports the empirical rejection rates (over 1000 replications) for both the McLeod type test \eqref{test_1} and the sample-splitting version \eqref{test_2}, which are denoted by $Q_n$ and $Q_n^S$, respectively. The maximum lag order is chosen to be $K\in \{3,6,9\}$. 

From the table, we find that, the McLeod type test $Q_n$ exhibits certain size distortion, especially for small samples and larger lag orders. Although the distortion decreases as the sample size increases, $Q_n$ remains oversized even at $n=400.$ In contrast, the sample-splitting statistic $Q_n^S$ delivers rejection probabilities much closer to the nominal level across all configurations.
Its size performance is stable for both positive and negative values of $\alpha_0$ and it remains well-controlled even when  $K$ is large.
Overall, the results confirm that sample splitting effectively mitigates estimation effects and yields a far more reliable portmanteau test in finite samples.

\begin{table}[htbp]
  \centering
  \caption{Size performance.}
  \footnotesize
  \label{tab_size}
  \begin{tabular}{@{} l l ccc ccc @{}}
    \toprule
    \multirow{2}{*}{$K$} & \multirow{2}{*}{Method} 
    & \multicolumn{3}{c}{$\alpha=0.2$} & \multicolumn{3}{c}{$\alpha=0.5$} \\
    \cmidrule(lr){3-5} \cmidrule(lr){6-8}
    & & $n=100$ & $n=200$ & $n=400$ & $n=100$ & $n=200$ & $n=400$ \\
    \midrule
    \multirow{2}{*}{3} 
    & $Q_n$       & 0.274 & 0.149 & 0.100 & 0.085 & 0.066 & 0.060 \\
    & $Q_n^S$ & 0.083 & 0.046 & 0.059 & 0.083 & 0.056 & 0.059 \\
    \addlinespace
    \multirow{2}{*}{6} 
    & $Q_n$       & 0.298 & 0.131 & 0.098 & 0.120 & 0.069 & 0.063 \\
    & $Q_n^S$ & 0.109 & 0.061 & 0.060 & 0.114 & 0.064 & 0.053 \\
    \addlinespace
    \multirow{2}{*}{9} 
    & $Q_n$      & 0.344 & 0.147 & 0.087 & 0.182 & 0.092 & 0.050 \\
    & $Q_n^S$ & 0.129 & 0.074 & 0.060 & 0.141 & 0.073 & 0.061 \\
    \midrule[\heavyrulewidth]
    \multirow{2}{*}{$K$} & \multirow{2}{*}{Method} 
    & \multicolumn{3}{c}{$\alpha=-0.2$} & \multicolumn{3}{c}{$\alpha=-0.4$} \\
    \cmidrule(lr){3-5} \cmidrule(lr){6-8}
    & & $n=100$ & $n=200$ & $n=400$ & $n=100$ & $n=200$ & $n=400$ \\
    \midrule
    \multirow{2}{*}{3} 
    & $Q_n$       & 0.253 & 0.142 & 0.112 & 0.124 & 0.095 & 0.082 \\
    & $Q_n^S$ & 0.087 & 0.058 & 0.057 & 0.095 & 0.075 & 0.071 \\
    \addlinespace
    \multirow{2}{*}{6} 
    & $Q_n$       & 0.235 & 0.101 & 0.100 & 0.123 & 0.075 & 0.076 \\
    & $Q_n^S$ & 0.107 & 0.068 & 0.055 & 0.110 & 0.082 & 0.073 \\
    \addlinespace
    \multirow{2}{*}{9} 
    & $Q_n$       & 0.218 & 0.105 & 0.074 & 0.120 & 0.066 & 0.062 \\
    & $Q_n^S$ & 0.131 & 0.073 & 0.063 & 0.127 & 0.090 & 0.077 \\
    \bottomrule
  \end{tabular}
\end{table}

We next evaluate the power of the proposed tests. 
To this end, we generate data from the ATM(2) model of \citet{zhu2023autoregressive}:
\[
T_i = T_{\epsilon_i} \circ [\alpha_1 \odot T_{i-1}] \circ [\alpha_2 \odot T_{i-2}],
\]
where the distortion maps are generated as in \eqref{eq:distortion}. 
We consider four parameter configurations, 
$(\alpha_1,\alpha_2)=\pm(0.2,0.1)$ and $(\alpha_1,\alpha_2)=\pm(0.5,0.2)$.
The model is then estimated as if the data were generated from an ATM(1) process, and the empirical rejection rates are reported in Table \ref{tab_power}.
\begin{table}[!htp]
  \centering
  \caption{Power performance.}
  \label{tab_power}
  \footnotesize
  \begin{tabular}{@{} l l ccc ccc @{}}
    \toprule
    \multirow{2}{*}{$K$} & \multirow{2}{*}{Method} 
    & \multicolumn{3}{c}{$(\alpha_1,\alpha_2)=(0.2,0.1)$} & \multicolumn{3}{c}{$(\alpha_1,\alpha_2)=(0.5,0.2)$} \\
    \cmidrule(lr){3-5} \cmidrule(lr){6-8}
    & & $n=100$ & $n=200$ & $n=400$ & $n=100$ & $n=200$ & $n=400$ \\
    \midrule
    \multirow{2}{*}{3} 
    & $Q_n$       & 0.295 & 0.306 & 0.460 & 0.861 & 0.962 & 0.997 \\
    & $Q_n^S$ & 0.818 & 0.863 & 0.933 & 0.910 & 0.976 & 0.999 \\
    \addlinespace
    \multirow{2}{*}{6} 
    & $Q_n$       & 0.308 & 0.276 & 0.404 & 0.966 & 0.990 & 0.998 \\
    & $Q_n^S$ & 0.960 & 0.966 & 0.975 & 0.972 & 0.996 & 0.999 \\
    \addlinespace
    \multirow{2}{*}{9} 
    & $Q_n$       & 0.239 & 0.258 & 0.364 & 0.984 & 0.997 & 0.999 \\
    & $Q_n^S$ & 0.985 & 0.988 & 0.995 & 0.987 & 0.997 & 0.997 \\
    \midrule[\heavyrulewidth]
    \multirow{2}{*}{$K$} & \multirow{2}{*}{Method} 
    & \multicolumn{3}{c}{$(\alpha_1,\alpha_2)=-(0.2,0.1)$} & \multicolumn{3}{c}{$(\alpha_1,\alpha_2)=-(0.5,0.2)$} \\
    \cmidrule(lr){3-5} \cmidrule(lr){6-8}
    & & $n=100$ & $n=200$ & $n=400$ & $n=100$ & $n=200$ & $n=400$ \\
    \midrule
    \multirow{2}{*}{3} 
    & $Q_n$       & 0.305 & 0.377 & 0.548 & 0.872 & 0.954 & 0.997 \\
    & $Q_n^S$ & 0.833 & 0.902 & 0.937 & 0.927 & 0.970 & 0.997 \\
    \addlinespace
    \multirow{2}{*}{6} 
    & $Q_n$       & 0.277 & 0.303 & 0.445 & 0.957 & 0.989 & 0.999 \\
    & $Q_n^S$ & 0.959 & 0.968 & 0.981 & 0.970 & 0.994 & 1.000 \\
    \addlinespace
    \multirow{2}{*}{9} 
    & $Q_n$       & 0.323 & 0.261 & 0.372 & 0.988 & 0.998 & 1.000 \\
    & $Q_n^S$ & 0.986 & 0.995 & 0.997 & 0.993 & 0.999 & 1.000 \\
    \bottomrule
  \end{tabular}
\end{table}

From the table, we see that both tests exhibit increasing power as the sample size grows and as the departure from the ATM(1) null model becomes more pronounced. Moreover, the split version $Q_n^S$ achieves substantially higher power than $Q_n$, especially under the weaker alternative $(\alpha_1,\alpha_2)=(0.2,0.1)$ and for smaller sample sizes. For the stronger alternative $(0.5,0.2)$, both tests approach full power, but $Q_n^S$ continues to outperform or match $Q_n$ across all configurations. The results are similar for negative values.

Overall, these findings demonstrate that sample splitting not only improves size control but also delivers superior power under model mis-specification, making the sample-splitting based test a more effective diagnostic tool in practice.

\section{Illustrative Data Analysis}\label{sec:app} 
 In this section, we analyze the distributional data of summer  temperatures from three major U.S. airports: Ted Stevens Anchorage International Airport (Alaska), San Francisco International Airport (California), and John F. Kennedy International Airport (New York). These locations represent distinct climate regimes—subarctic, cool-summer Mediterranean, and humid subtropical—providing a diverse set of distributional dynamics for evaluating the ATM model. This dataset has been analyzed in both \cite{zhu2023autoregressive} and \cite{ghodrati2023on}. 
 
\begin{figure}[!htbp]
  \centering
  \begin{subfigure}[b]{0.3\textwidth} 
    \includegraphics[width=\textwidth]{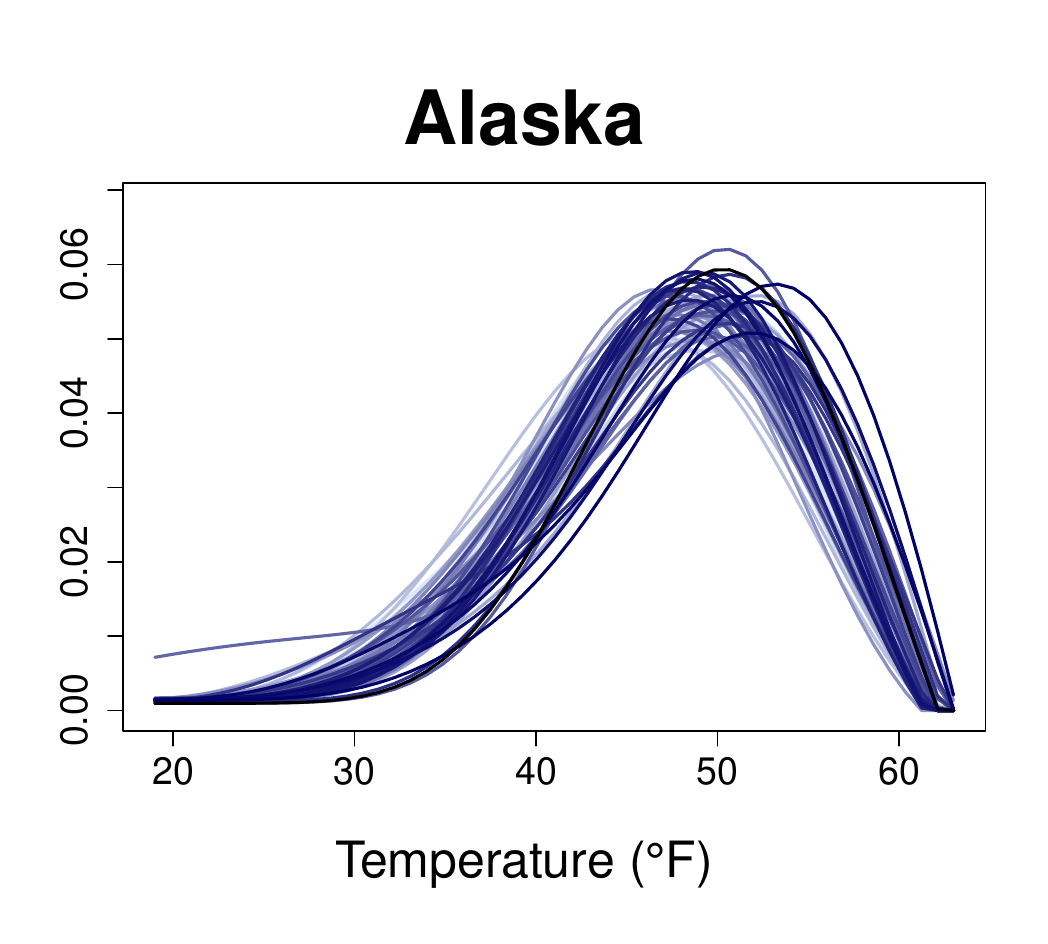} 
    \label{fig:alaskadis}
  \end{subfigure}
  \hfill 
  \begin{subfigure}[b]{0.3\textwidth}
    \includegraphics[width=\textwidth]{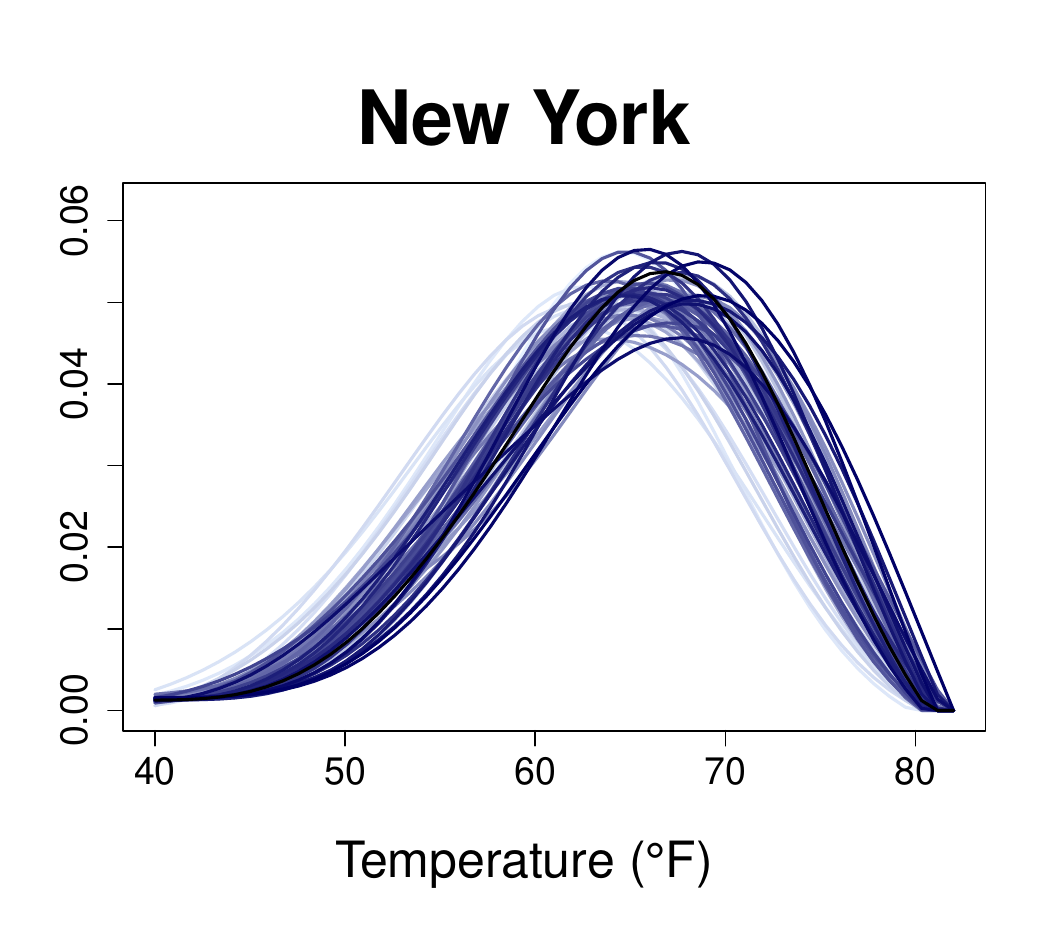}
    \label{fig:newyorkdis}
  \end{subfigure}
  \hfill
  \begin{subfigure}[b]{0.3\textwidth}
    \includegraphics[width=\textwidth]{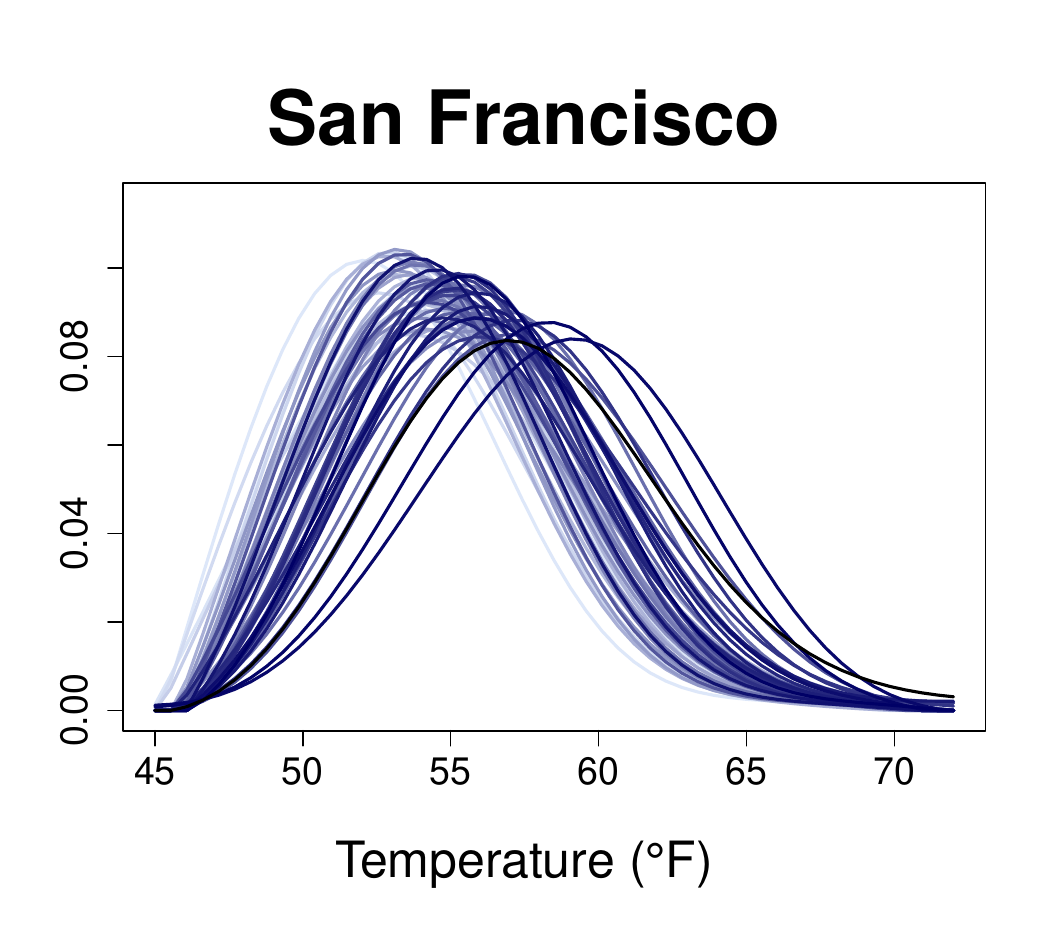}
    \label{fig:sanfranciscodis}
  \end{subfigure}
  \caption{Time series of  distributions of daily minimum temperatures during summer (June–September) for 1960–2019 at three U.S. airports: Ted Stevens Anchorage (Alaska), San Francisco (California), and John F. Kennedy (New York). The color gradient of the distribution curves encodes the temporal order, with lighter shades denoting earlier years.} 
  \label{fig:distributions}
\end{figure}

For each airport, the dataset contains annual distributions of daily minimum temperatures from June 1 to September 30 for the years 1960–2018, see  Figure \ref{fig:distributions} for a visual illustration. We see that all three distribution series tend to shift toward higher temperatures over time, with earlier-year curves appearing mostly on the left and more recent curves progressively moving to the right. This systematic “rightward shift’’ can be naturally captured and quantified through the Wasserstein distance and the ATM modeling framework.

 As a preliminary step, we apply the proposed diagnostic tests to assess whether the temperature distributions at each location are compatible with an ATM(1) specification.  The p-values of both the McLeod type test and sample-splitting portmanteau test are reported in Table \ref{tab:pvalue}. At the 5\% significance level, both tests reject the ATM(1) specification for Alaska and San Francisco at all lags, whereas for New York the model is rejected only at lag
 $h=9$. This indicates that the ATM(1) model provides a reasonable approximation for the New York data.

\begin{table}[!htbp]
\centering
\small
\setlength{\tabcolsep}{10pt}
\renewcommand{\arraystretch}{1.3}
\caption{p-values (0 indicates values smaller than $10^{-3}$) for the McLeod type test and the sample-splitting portmanteau test.}
\label{tab:pvalue}
\begin{tabular}{lccc ccc}
\toprule
& \multicolumn{3}{c}{\textbf{McLeod Type}} 
& \multicolumn{3}{c}{\textbf{Sample-Splitting Type}} \\
\hline
\diagbox[width=9em,height=3em]{\textbf{Locations}}{\textbf{Lags}}
& \textbf{h=3} & \textbf{h=6} & \textbf{h=9}
& \textbf{h=3} & \textbf{h=6} & \textbf{h=9} \\
\midrule
\textbf{Alaska}        & 0     & 0     & 0     & 0     & 0     & 0     \\
\textbf{New York}      & 0.178 & 0.086 & 0.003 & 0.460 & 0.343 & 0.029 \\
\textbf{San Francisco} & 0.021 & 0     & 0     & 0     & 0     & 0     \\
\bottomrule
\end{tabular}
\end{table}

\begin{figure}[!htbp]
  \centering
  
  \begin{subfigure}[b]{0.3\textwidth} 
    \includegraphics[width=\textwidth]{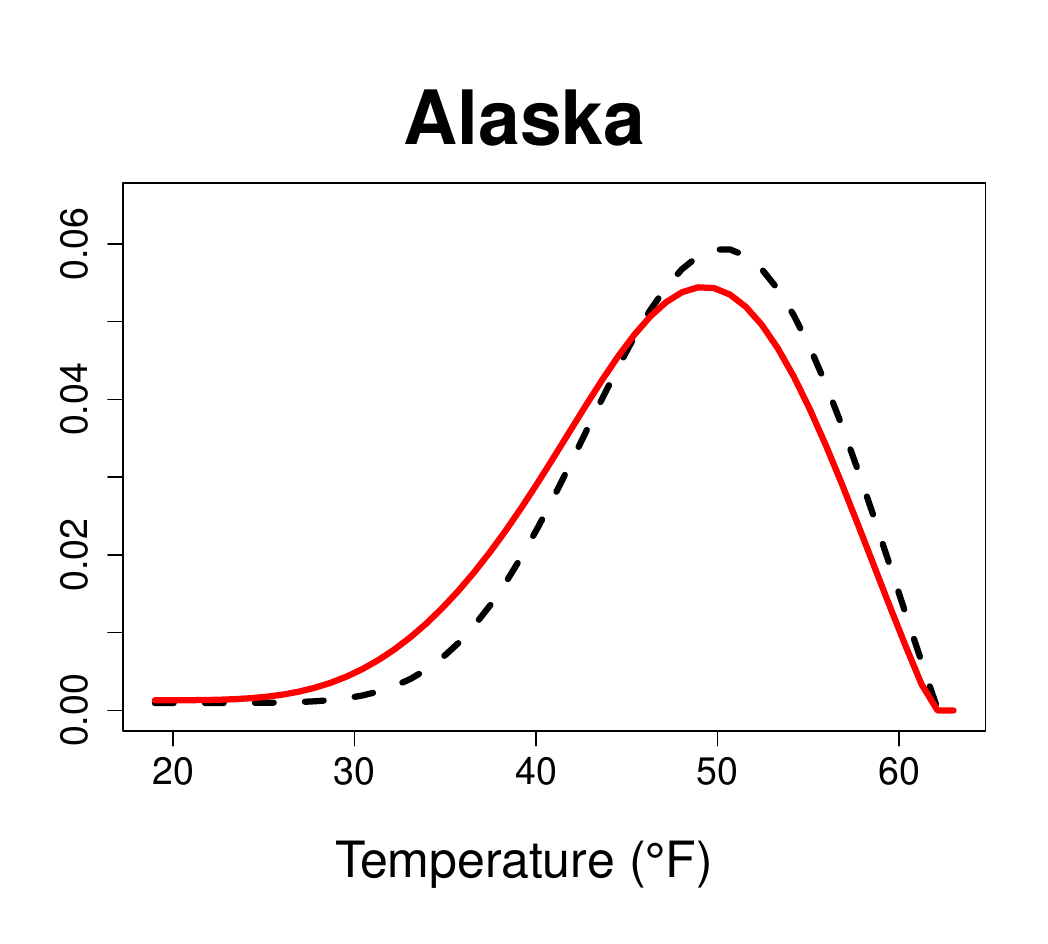} 
    \label{fig:alaskapred}
  \end{subfigure}
  \hfill 
  \begin{subfigure}[b]{0.3\textwidth}
    \includegraphics[width=\textwidth]{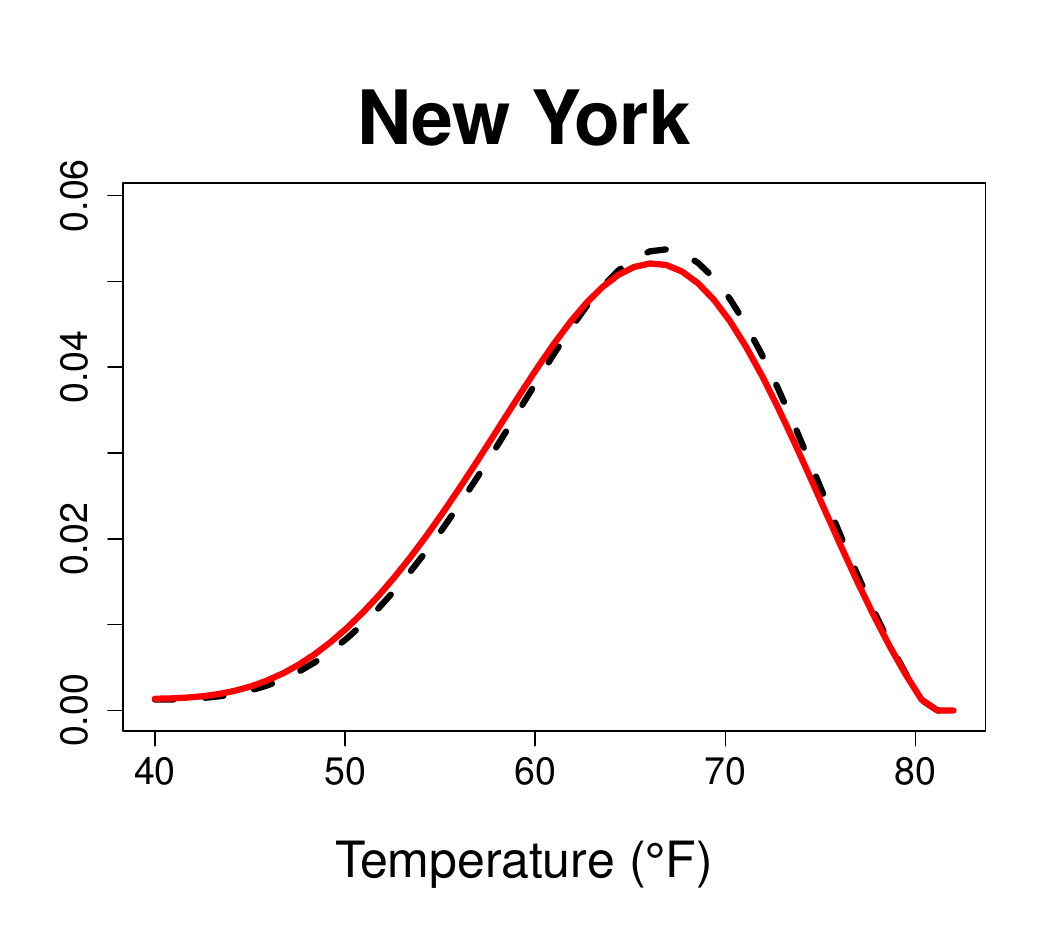}
    \label{fig:newyorkpred}
  \end{subfigure}
  \hfill
  \begin{subfigure}[b]{0.3\textwidth}
    \includegraphics[width=\textwidth]{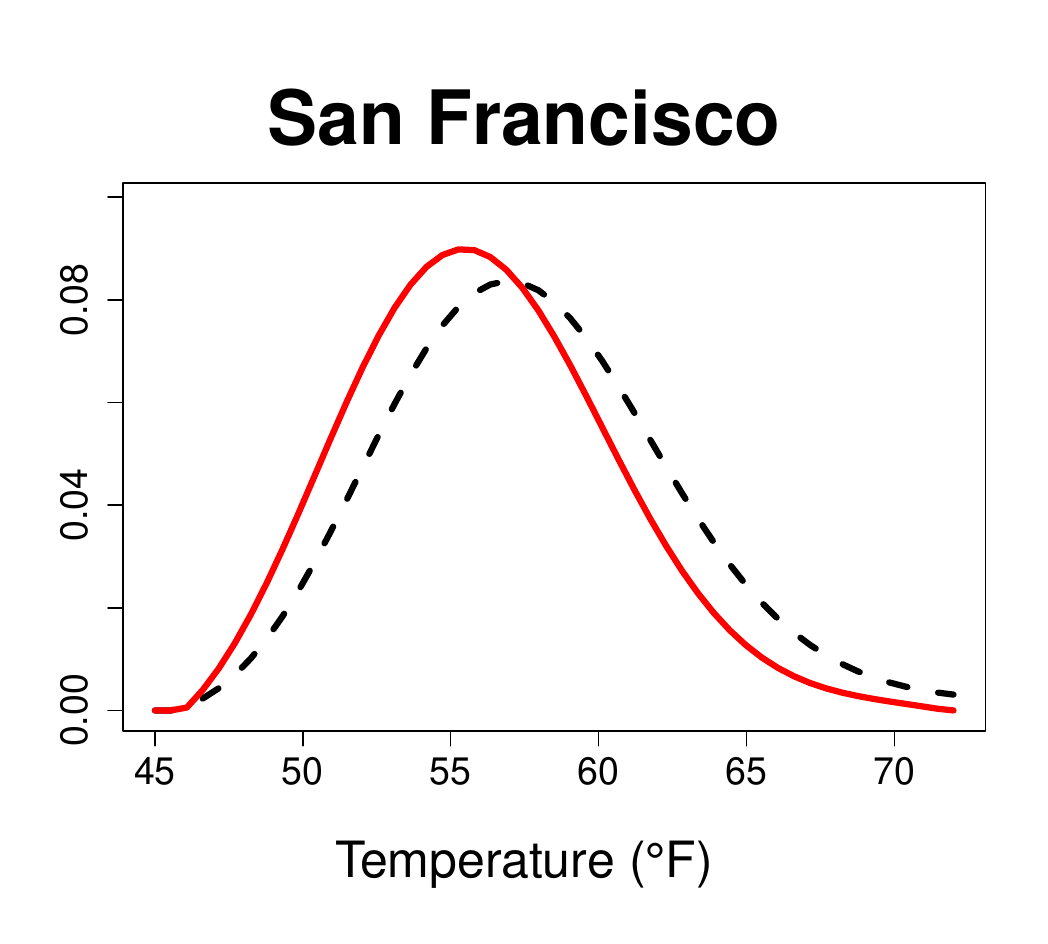}
    \label{fig:sanfranciscopred}
  \end{subfigure}
  \caption{Model validation for the year 2018. The observed distribution of daily minimum summer temperatures is shown in black, compared against the predicted distribution from the fitted ATM(1) model (red) for each airport.} 
  \label{fig:predictions}
\end{figure}

 We further corroborate these findings by fitting the ATM(1) model to each dataset and estimating the fitted distribution for the year 2018, see  Figure \ref{fig:predictions}. Visually, the prediction for Alaska exhibits the largest discrepancy, with the fitted curve deviating noticeably from the observed distribution, whereas the prediction for New York aligns most closely with the data.  

Finally, using a 50-year training period (1960–2009) and rolling validation windows from 1968 to 2017, we predicted the distributions for the years 2010–2018 and computed the corresponding Wasserstein distances between the predicted and observed distributions. The results are summarized in Table \ref{tab:predict}. Consistent with our earlier diagnostic findings, Alaska exhibits the largest prediction error, while New York shows the smallest, indicating stronger conformity with the ATM(1) model.

\begin{table}[!htbp]
\centering
\caption{Wasserstein distances between predicted and observed distributions.}
\label{tab:predict}
\begin{tabular}{lccc}
\toprule
\textbf{Year} & \textbf{Alaska} & \textbf{New York} & \textbf{San Francisco} \\
\midrule
2010 & 0.7197 & 1.4122 & 1.1353 \\
2011 & 0.5619 & 1.0701 & 0.1159 \\
2012 & 0.4439 & 0.7529 & 0.5287 \\
2013 & 2.0531 & 0.5012 & 0.9881 \\
2014 & 1.1575 & 0.4089 & 2.6707 \\
2015 & 1.2996 & 1.6848 & 1.9061 \\
2016 & 2.8522 & 1.6503 & 0.3621 \\
2017 & 1.5246 & 0.3763 & 1.5630 \\
2018 & 1.6527 & 0.8876 & 0.1376 \\
\midrule
\textbf{average} & 1.3628 & 0.9716 & 1.0453 \\
\bottomrule
\end{tabular}
\end{table}

\clearpage

\bibliographystyle{apalike}
\bibliography{reference}

@article {davis2024sample,
    AUTHOR = {Davis, Richard A. and Fernandes, Leon},
     TITLE = {Sample splitting and assessing goodness-of-fit of time series},
   JOURNAL = {Biometrika},
  FJOURNAL = {Biometrika},
    VOLUME = {112},
      YEAR = {2025},
    NUMBER = {2},
     PAGES = {Paper No. asaf017, 21},
      ISSN = {0006-3444,1464-3510},
   MRCLASS = {62M10 (62G10)},
  MRNUMBER = {4929421},
       DOI = {10.1093/biomet/asaf017},
       URL = {https://doi.org/10.1093/biomet/asaf017},
}

@article{mazzuco2015,
	author = {Mazzuco, Stefano and Scarpa, Bruno},
	title = {Fitting age-specific fertility rates by a flexible generalized skew normal probability density function},
	journal = {Journal of the Royal Statistical Society Series A: Statistics in Society},
	volume = {178},
	number = {1},
	pages = {187-203},
	year = {2015}
}

@article{shang2017,
	author = {Han Lin Shang and Rob J. Hyndman},
	title = {Grouped Functional Time Series Forecasting: an Application to Age-Specific Mortality Rates},
	journal = {Journal of Computational and Graphical Statistics},
	volume = {26},
	number = {2},
	pages = {330-343},
	year  = {2017},
	publisher = {Taylor & Francis},
}

@article{dubey2020frechet,
	title = {Fr{\'e}chet change-point detection},
	author = {Dubey, Paromita and Müller, Hans-Georg},
	journal = {The Annals of Statistics},
	volume = {48},
	number = {6},
	pages = {3312--3335},
	year = {2020},
	publisher = {Institute of Mathematical Statistics}
}

@article{jiang2023two,
	title={Two-Sample and Change-Point Inference for Non-{E}uclidean Valued Time Series},
	author={Jiang, Feiyu and Zhu, Changbo and Shao, Xiaofeng},
	journal={Electronical Journal of Statistics},
    volume={18},
    number={1},
   page={848-894},
	year={2024}
}

@article{zhu2023autoregressive,
  title={Autoregressive optimal transport models},
  author={Zhu, Changbo and M{\"u}ller, Hans-Georg},
  journal={Journal of the Royal Statistical Society Series B: Statistical Methodology},
  volume={85},
  number={3},
  pages={1012--1033},
  year={2023},
  publisher={Oxford University Press US}
}

@article{ghodrati2023on,
  title={On distributional autoregression and iterated transportation},
  author={Ghodrati, Laya and Panaretos, Victor M},
  journal={Journal of Time Series Analysis},
  volume={45},
  number={5},
  pages={739--770},
  year={2024},
  publisher={Wiley Online Library}
}

@article{zhang2022wasserstein,
  title={Wasserstein autoregressive models for density time series},
  author={Zhang, Chao and Kokoszka, Piotr and Petersen, Alexander},
  journal={Journal of Time Series Analysis},
  volume={43},
  number={1},
  pages={30--52},
  year={2022},
  publisher={Wiley Online Library}
}

@article{wu2004limit,
  title={Limit theorems for iterated random functions},
  author={Wu, Wei Biao and Shao, Xiaofeng},
  journal={Journal of Applied Probability},
  volume={41},
  number={2},
  pages={425--436},
  year={2004},
  publisher={Cambridge University Press}
}

@article{jiang2024testing,
  title={Testing serial independence of object-valued time series},
  author={Jiang, Feiyu and Gao, Hanjia and Shao, Xiaofeng},
  journal={Biometrika},
  volume={111},
  number={3},
  pages={925--944},
  year={2024},
  publisher={Oxford University Press}
}

@article{zhangshao2023,
  title={Another look at bandwidth-free inference: A sample splitting approach},
  author={Zhang, Yi and Shao, Xiaofeng},
  journal={Journal of the Royal Statistical Society Series B: Statistical Methodology},
  volume={86},
 number={1},
pages = {246--272},
  year={2024},
  publisher={Oxford University Press US}
}

@article {gaowangshao2023,
    AUTHOR = {Gao, Hanjia and Wang, Runmin and Shao, Xiaofeng},
     TITLE = {Dimension-agnostic change point detection},
   JOURNAL = {Journal of Econometrics},
  FJOURNAL = {Journal of Econometrics},
    VOLUME = {250},
      YEAR = {2025},
     PAGES = {106012},
      ISSN = {0304-4076,1872-6895},
   MRCLASS = {99-01},
  MRNUMBER = {4903388},
       DOI = {10.1016/j.jeconom.2025.106012},
}

@article{jiang2022,
title={Wasserstein multivariate auto-regressive models for modeling distributional time series and its application in graph learning},
author={Yiye Jiang},
journal={arXiv:2207.05442},
year={2022}
}

@article{chen2023,
  title={Wasserstein regression},
  author={ Yaqing Chen and Zhenhua Lin and Hans-Georg Müller},
  journal={Journal of the American Statistical Association},
volume = {118},
number = {542},
pages = {869-882},
  year={2023}
}

@article {Chang21,
    AUTHOR = {Chang, Jinyuan and Cheng, Guanghui and Yao, Qiwei},
     TITLE = {Testing for unit roots based on sample autocovariances},
   JOURNAL = {Biometrika},
  FJOURNAL = {Biometrika},
    VOLUME = {109},
      YEAR = {2022},
    NUMBER = {2},
     PAGES = {543--550},
      ISSN = {0006-3444,1464-3510},
   MRCLASS = {62M10},
  MRNUMBER = {4430974},
       DOI = {10.1093/biomet/asab034},
       URL = {https://doi.org/10.1093/biomet/asab034},
}

@article{Lunde19,
	author = {R. Lunde},
	title = {Sample splitting and weak assumption inference for time series},
	year = {2019},
	journal = {https://arxiv.org/abs/1902.07425}
}

@article{zhangzhushao2025,
  title={Change-point detection for object-valued time series},
  author={Yi Zhang and Changbo Zhu and Xiaofeng Shao},
  journal={Journal of Business and Economic Statistics, to appear},
  year={2025}
}

@article{zhumuller2024JoE,
	title={Spherical autoregressive models, with application to distributional and compositional time series},
	author={Zhu, Changbo and M\"uller, Hans-Georg"},
	journal={Journal of Econometrics},
	volume={239},
	pages={105389},
	year={2024}
}

@book{Li2003Diagnostic,
  author    = {Wai Keung Li},
  title     = {Diagnostic Checks in Time Series},
  year      = {2003},
  publisher = {Chapman \& Hall/CRC}}

@book{hall2014martingale,
  title={Martingale Limit Theory and Its Application},
  author={Hall, Peter and Heyde, Christopher C},
  year={2014},
  publisher={Academic press}
}

@article{li1981distribution,
  title={Distribution of the residual autocorrelations in multivariate ARMA time series models},
  author={Li, Wai Keung and McLeod, A Ian},
  journal={Journal of the Royal Statistical Society Series B: Statistical Methodology},
  volume={43},
  number={2},
  pages={231--239},
  year={1981},
  publisher={Oxford University Press}
}

@article{ljung1978measure,
  title={On a measure of lack of fit in time series models},
  author={Ljung, Greta M and Box, George EP},
  journal={Biometrika},
  volume={65},
  number={2},
  pages={297--303},
  year={1978},
  publisher={Oxford University Press}
}

@article{li1994squared,
  title={On the squared residual autocorrelations in non-linear time series with conditional heteroskedasticity},
  author={Li, Wai Keung and Mak, Tuck Kiong},
  journal={Journal of Time Series Analysis},
  volume={15},
  number={6},
  pages={627--636},
  year={1994},
  publisher={Wiley Online Library}
}

@article{mcleod1978distribution,
  title={On the distribution of residual autocorrelations in Box--Jenkins models},
  author={McLeod, AI},
  journal={Journal of the Royal Statistical Society: Series B (Methodological)},
  volume={40},
  number={3},
  pages={296--302},
  year={1978},
  publisher={Wiley Online Library}
}

@article{box1970distribution,
  title={Distribution of residual autocorrelations in autoregressive-integrated moving average time series models},
  author={Box, George EP and Pierce, David A},
  journal={Journal of the American statistical Association},
  volume={65},
  number={332},
  pages={1509--1526},
  year={1970},
  publisher={Taylor \& Francis}
}

@article{newbold1980equivalence,
  title={The equivalence of two tests of time series model adequacy},
  author={Newbold, Paul},
  journal={Biometrika},
  volume={67},
  number={2},
  pages={463--465},
  year={1980},
  publisher={Oxford University Press}
}

@article{ansley1979finite,
  title={On the finite sample distribution of residual autocorrelations in autoregressive-moving average models},
  author={Ansley, Craig F and Newbold, Paul},
  journal={Biometrika},
  volume={66},
  number={3},
  pages={547--553},
  year={1979},
  publisher={Oxford University Press}
}

@article{li1992asymptotic,
  title={On the asymptotic standard errors of residual autocorrelations in nonlinear time series modelling},
  author={Li, Wai Keung},
  journal={Biometrika},
  volume={79},
  number={2},
  pages={435--437},
  year={1992},
  publisher={Oxford University Press}
}

\end{document}